\documentclass[preprint,12pt,numbers]{elsarticle}
\usepackage{graphicx}
\usepackage{graphicx}
\usepackage{multirow}
\usepackage{xcolor}
\usepackage{amsmath}
\usepackage{amsfonts}
\usepackage{amssymb}
\usepackage{hyperref}
\usepackage{subfigure}
\usepackage{array}

\biboptions{numbers,sort&compress}
\journal{Journal of Subatomic particle and Cosmology}
\begin{document}
\begin{frontmatter}
\title{3+1 neutrino mixings model with $A_4$ triplet Majorana neutrino} 
%
\author[label1]{Mayengbam Kishan Singh} 
\ead{kishanmayengbam@gmail.com}
\affiliation[label1]{organization={Department of Physics, D.M. College of Science, Dhanamanjuri University},
            city={Imphal},
            postcode={795001}, 
            state={Manipur},
            country={India}}
\author[label2,label3]{N. Nimai Singh} 
\ead{nimai03@yahoo.com}
\affiliation[label2]{organization={Department of Physics,  Manipur University}, 
            city={Imphal},
            postcode={795003}, 
            state={Manipur},
            country={India}}
\affiliation[label3]{organization={Research Institute of Science and Technology (RIST)},
            city={Imphal},
            postcode={795003}, 
            state={Manipur},
            country={India}}
\begin{abstract}
We study a 3+1 active-sterile neutrino mixings model using an $A_4$ triplet right-handed neutrino $\nu_R$ and a singlet eV-scale sterile neutrino under $A_4\times Z_3 \times Z_2$ discrete symmetry.  Four scalar flavons are considered to reproduce neutrino oscillation parameters within the experimental 3$\sigma$ range. The model also studies the effective mass parameter in neutrinoless double beta decay experiments. Deviation from $\mu-\tau$ symmetry in the active neutrino mass matrix is generated through an antisymmetric interaction of $\nu_R$. This model successfully explains active-sterile neutrino mixings consistent with the cosmological upper bound on the sum of active neutrino mass $\sum m_i < 0.113$ eV (0.145 eV) in NH(IH).
\end{abstract}
\end{frontmatter}
\section{Introduction}
\label{section1}
With the successful detection of the Higgs boson at the LHC, the Standard Model (SM) of particle physics is standing tall as the strongest theory in high energy physics. However, unanswered puzzles still require understanding Beyond Standard Model (BSM) theories. Among them, one of the most exciting questions is the origin of neutrino masses and their mixings among different flavors. The extension of SM gauge symmetry with certain non-Abelian discrete symmetries such as $S_3, S_4, A_4, A_5, \Delta(27),$ etc. has its strengths in describing the neutrino problem. Consequently, neutrino masses and mixings models based on $A_4$\cite{altarelli2010discrete,king2007a4,puyam2022deviation}, $S_4$\cite{ma2006neutrino,altarelli2009revisiting,vien2022lepton}, $S_3$\cite{mohapatra2006s3,Vien:2021uxw,koide2007s}, $A_5$ \cite{ding2019neutrino,puyam2025a5,novichkov2019modular}, $\Delta (27)$\cite{hernandez20163,wilina2025perturbation,wilina2024deviations}, etc. have been widely studied. Significant observations of anomalies in the LSND \cite{aguilar2001evidence}, MiniBooNE \cite{aguilar2018significant,2021microboone}, Gallium anti-neutrino and reactor anti-neutrino experiments motivated the need for extension of the three neutrino paradigm to a 3+1  \cite{Zhang2011,mksingh,das2019active,mksingh2023,vien2022b,vien2021,singh2024modular}, 3+2 \cite{donini2012minimal,babu2016light,karagiorgi2007leptonic}, 3+1+1  \cite{kuflik2012neutrino,huang2013mev,nelson2011effects} schemes. The 3+1 active-sterile neutrino mixing theory is the most successful scenario in describing mixings between active and light sterile neutrinos. A new state of neutrino called $sterile$ neutrino in the three neutrino paradigm has been proposed \cite{Zhang2011} in a mechanism called the minimal extended seesaw (MES). In this seesaw, the active neutrinos are generated at the sub-eV scale, while heavier sterile neutrinos at the eV or KeV scale are naturally produced, and the lightest neutrino mass is zero. The current neutrino oscillation data still allows the case where the lightest neutrino mass $m_{lightest} \equiv m_{1} (m_{3})$ vanishes. A recent cosmological probe on the upper bound of the sum of neutrino mass gives $\sum m_i \lesssim 0.113\ (0.145)$ eV for NH (IH)\cite{adame2025desi}. \\

 Another important aspect of neutrino physics is the exact nature of neutrinos: Dirac or Majorana particle. The Majorana nature of neutrinos can be probed through a beta decay process called the neutrinoless double beta decay ($ \nu 0\beta\beta$). In this work, we consider the $A_4$ discrete symmetry with four scalars $\phi_1, \phi_2, \chi $ and $\zeta$ to explain neutrino phenomenologies by considering active-sterile mixings in 3+1 scheme where sterile neutrino mass is in eV scale. In earlier studies of the MES mechanism under $A_4$, the heavy right-handed neutrinos are taken as singlets, which requires an additional scalar to generate neutrino mass structures consistent with data. The present work is different and advantageous from others in that we consider the heavy right-handed neutrino $\nu_R$ as triplet under $A_4$ in the MES mechanism \cite{Zhang2011} with only four scalars. In addition,  deviation from the $\mu-\tau$ symmetry of neutrino mass matrix is produced by the antisymmetric product of $\nu_R$ with $\phi_2$. This article is organized as follows: the next section \ref{model} gives a detailed description of the structure of the model, followed by the numerical analysis in section \ref{numerical}. The numerical results are given in section \ref{result}. We conclude this article with a conclusion and discussion in section \ref{conclusion}.
\section{Description of the model}
\label{model}
\begin{table}
\centering
\small
\begin{tabular}{ccccc}
\hline
$Fields$ & $SU(2)_L\times U(1)_Y$ & $A_4$ & $Z_3$ & $Z_2$  \\
\hline
$l$& (2,$-1/2$) & 3  & 1 & + \\
$e_R, \mu_R, \tau_R $& (1,$-1$) &1,1$^{\prime\prime}$,1$^{\prime}$ &$\omega$ & + \\
H & (2,+1/2) & 1 & 1 & +  \\
$\nu_R$ & (1,0) & 3 & $\omega$ & $-$ \\
S & (1,0) & 1 & $\omega$ & $-$ \\
$\phi_1$ & (1,0) & 3 & $\omega^2$ & + \\
$\phi_2$ &(1,0) & 3 & $\omega^2$ & $-$ \\
$\chi$ & (1,0) & 1 & $\omega^2$ & $-$ \\
$\zeta$ & (1,0) & 3 & 1 & + \\
\hline
\end{tabular}
\caption{\footnotesize Particle contents of the model and their group representations and corresponding hypercharges. }
\label{table1}
\end{table}
The particle contents of the model and their group representations and hypercharges under $SU(2)_L\times U(1)_Y \times A_4\times Z_3\times Z_2$ are given in Table \ref{table1}. The model uses the Weinberg dimension 5 operator to generate lepton mass matrices. The scalar $\phi_1$ generates a diagonal charged lepton mass matrix $M_L$ after spontaneous symmetry breaking while the symmetric and antisymmetric interactions of $\phi_2$ with the left-handed lepton doublet $\ell$ and right-handed neutrino $\nu_R$ generate the Dirac neutrino mass matrix $M_D$ defined by the Yukawa constants $y_1,y_2$ and $y_3$. Finally, the triplet scalar $\zeta$ is responsible for the sterile neutrino mass matrix $M_S$. The scale of the VEV of $\zeta$ will fix the mass of the sterile neutrino developed in this MES mechanism \footnote{For instance, $M_S\sim \mathcal{O}(10^2)$GeV will give sterile neutrinos in eV scale. In contrast, $M_S\sim \mathcal{O}(10)$TeV will generate KeV scale sterile neutrino mass}. The invariant interaction terms in the Lagrangian based on particle contents and group charges given in Table \ref{table1} are
\begin{align}
- &\mathcal{L} = \frac{y_e}{\Lambda}(\bar{l}H\phi_1)_1e_R+\frac{y_{\mu}}{\Lambda}(\bar{l}H\phi_1)_{1^{\prime}}\mu_R+\frac{y_{\tau}}{\Lambda}(\bar{l}H\phi_1)_{1^{\prime\prime}}\tau_R+\frac{y_1}{\Lambda}(\bar{l}\tilde{H}\phi_2)_a\nu_R  \nonumber \\
 +\ & \frac{y_2}{\Lambda}(\bar{l}\tilde{H}\phi_2)_s\nu_R 
+ \frac{y_3}{\Lambda}(\bar{l}\tilde{H}\nu_R)_1\chi+\frac{y_R}{2\Lambda}\left(\chi^2\bar{\nu_R}^c\nu_R + \phi_2 \chi\bar{\nu_R}^c\nu_R\right) + \frac{y_s}{2}\bar{S}^c\zeta\nu_R + h.c.
\end{align}
Here, the subscripts $a$ and $s$ represent the antisymmetric and symmetric tensor products of triplets in $A_4$. Choosing the most general VEV alignments along $\langle \phi_1\rangle = (v,0,0), \langle \phi_2\rangle = (v,v,v),\langle \chi\rangle = v, \langle \zeta\rangle = (v_{\zeta},0,0),$  we get a diagonal $M_L$ given by 
\begin{equation}
M_L = \frac{\langle H \rangle v}{\Lambda}\left(
\begin{array}{ccc}
 y_e & 0 & 0 \\
 0 & y_{\mu} & 0 \\
 0 & 0 & y_{\tau} \\
\end{array}
\right).
\end{equation}
The charged lepton masses $m_{e}, m_{\mu}$ and $m_{\tau}$ are obtained with proper choices of the Yukawa coupling constants $y_e, y_{\mu}$ and $y_{\tau}$ respectively. Similarly, the Dirac neutrino mass matrix $M_D$, the heavy right-handed Majorana mass matrix $M_R$ and sterile neutrino mass matrix $M_S$ are respectively given by 
\begin{align}
M_D = & \frac{\langle H \rangle v}{\Lambda}\left(
\begin{array}{ccc}
 2 y_1 + y_3 & -y_1-y_2 & y_2-y_1 \\
 y_2-y_1 & 2 y_1 & - y_1-y_2+y_3 \\
 -y_1-y_2 & -y_1+y_2+y_3 & 2 y_1 \\
\end{array}
\right), \\
M_R = & \frac{v^2}{2\Lambda}\left(
\begin{array}{ccc}
 y_R & 0 & 0 \\
 0 & 0 & y_R \\
 0 & y_R & 0 \\
\end{array}
\right), \ \ \ \ \text{and }\ M_S = y_s v_{\zeta} \left(
\begin{array}{ccc}
 1 & 0 & 0 \\
\end{array}
\right).
\end{align}
Using these mass matrices we apply the MES mechanism and the resulting active and sterile neutrino mass matrices are given by 
\begin{align}
m_{\nu} = m_o \left(
\begin{array}{ccc}
m_{11} & m_{12} & m_{13} \\
m_{21} & m_{22} & m_{23} \\
m_{31} & m_{32} & m_{33} \\
\end{array}
\right), \ \ \  \text{and }\  \ m_s=(y_s v_{\zeta})^2/M.
\label{mv}
\end{align}
Here, \begin{align*}
m_{11} &= -2 (y_1-y_2) (y_1+y_2), \\
m_{12} &= m_{21} = y_1^2+y_1 (y_3-4y_2)+y_2 (y_3-y_2), \\
m_{13} &= m_{31} = y_1^2+y_1 (4y_2+y_3)-y_2 (y_2+y_3), \\
m_{22} &= 4 y_1 (y_1+y_2-y_3), \\
m_{23} &= m_{32} = -5 y_1^2+ 2 y_1 y_3 +y_2^2-y_3^2, \\
m_{33} &=  -4 y_1 (-y_1+y_2+y_3), \\
\text{and } M = \frac{ v^2 y_R }{2\Lambda}.
\end{align*}
The overall coefficient $m_o$ in Eq.(\ref{mv}) is a constant having the dimension of mass in eV scale. It is important to note that the additional symmetries $Z_3\times Z_2$ are imposed to avoid interactions such as $\frac{y_e}{\Lambda}(\bar{l}H\phi_2)_1e_R,\frac{y_e}{\Lambda}(\bar{l}H\phi_1)_1\nu_{R},\frac{y_e}{\Lambda}(\bar{l}H\zeta)_1\nu_R$, $\frac{y}{\Lambda}(\bar{l}H\phi_1)_1S$, etc. in the Lagrangian. We proceed to the numerical analysis of the model in the next section. 
\begin{table}
{\begin{tabular} {@{}ccc @{}}
\hline
\rule{0pt}{4ex} Parameter &	NH (best-fit$\pm 1\sigma$) &	IH (best-fit$\pm 1\sigma$)  \\
\hline
\rule{0pt}{4ex} $\Delta m^2_{21}: [10^{-5} eV^2]$ & 6.92 – 8.05 $(7.49^{+0.19}_{-0.19})$  & 6.92 – 8.05 $(7.49^{+0.19}_{-0.19})$ \\
$\vert\Delta m^2_{3\ell}\vert: [10^{-3} eV^2]$	& 2.463 – 2.606 $(2.534^{+0.025}_{-0.023})$  & 2.438 – 2.584 $(2.510^{+0.024}_{-0.025})$ \\
$\sin^2\theta_{12} $	& 0.275 – 0.345 $(0.307^{+0.012}_{-0.011})$ &  0.275 – 0.345 $(0.303^{+0.012}_{-0.011})$  \\
$\sin^2\theta_{23}$ & 0.430 – 0.596	$(0.561^{+0.012}_{-0.015})$  &0.437 – 0.597 $(0.562^{+0.012}_{-0.015})$ 		 \\
$\sin^2\theta_{13}/10^{-2}$ & 2.023 – 2.376	$(2.195^{+0.054}_{-0.058})$ & 2.053 – 2.397 $(2.2224^{+0.056}_{-0.057})$  \\
$\delta_{\rm CP}/^o$ &	96 - 422 $(177^{+42}_{-0.25})$	& 201 - 348 $(285^{+25}_{-28})$	 \\
$r=\sqrt{\frac{\Delta m_{21}^2}{\vert\Delta m_{3l}^2\vert}} $ & 0.1676 - 0.1757 (0.1719)  & 0.1684 - 0.1765 (0.1727)\\
$\vert U_{14}\vert^2 $  & 0.012 - 0.047 & 0.012 - 0.047 \\
$\vert U_{24}\vert^2 $  & 0.005 - 0.03 &  0.005 - 0.03 \\
$\vert U_{34}\vert^2 $  & 0 - 0.16 & 0 - 0.16 \\
\hline
\end{tabular}\label{nufitdata}}
\caption{Updated global-fit data for three neutrino oscillation, NuFIT 6.0(2024) \protect\cite{esteban2025nufit}. For 3+1 mixing, data is taken from  Refs. \protect\cite{barrylight,vien2022b,Gariazzo_2016}.}
\end{table}
\section{Numerical Analysis}
\label{numerical}
Since the charged lepton mass matrix is diagonal, its contribution to the lepton mixing matrix $U_L$ is an identity matrix. Then, the PMNS lepton mixing matrix is given by $U = U_{\nu}$. To diagonalise the active neutrino mass matrix $m_{\nu}$, we define a hermitian matrix $h = m_{\nu}^{\dagger} m_{\nu}$. The neutrino mixing matrix $U_{\nu}$ is given by the diagonalisation relation $D = U_{\nu}^{\dagger} m_{\nu} U_{\nu}$, where $D$ is a diagonal matrix having real eigenvalues $m_1^2, m_2^2, m_3^2$.  Here, $m_1, m_2, m_3$ are the active neutrino masses. The full $(4\times 4)$ active-sterile neutrino mixing matrix is given by  \cite{v441982}
\begin{equation}
V \simeq \left(\begin{array}{ccc}
(1-\frac{1}{2}RR^{\dagger})U & R \\ 
-R^{\dagger}U & 1-\frac{1}{2}R^{\dagger}R
\end{array} \right),
\label{V44}
\end{equation}
where $U$ represents the $3\times 3$ active neutrino mixing matrix and $R$ represents the strength of active-sterile mixing given by
\begin{eqnarray}
R=&M_DM_R^{-1}M_S^T(M_SM_R^{-1}M_S^T)^{-1}.
\label{RR}
\end{eqnarray}
Deviations from unitarity in $U$ due to the presence of sterile neutrino is defined by $\frac{1}{2}RR^{\dagger}$. The neutrino mixing matrix $V$ is  parametrized by six mixing angles $(\theta_{12},\ \theta_{13},\ \theta_{23},\ \theta_{14},\theta_{24},\ \theta_{34})$, three Dirac phases $(\delta_{CP},\ \delta_{14},\ \delta_{24})$ and three Majorana phases $(\alpha,\ \beta,\ \gamma)$ \cite{navas2024review,gariazzo2016light}. The mixing angles are calculated from the elements $V_{ij}$ of $V$ using the relations  
\begin{align} \label{C1anglessolve}
\sin^2\theta_{14}\ &=\ \vert V_{e4}\vert ^2,\ \  
\sin^2\theta_{24}\ =\ \frac{\vert V_{\mu 4}\vert ^2}{1-\vert V_{e4}\vert ^2},\ \ 
\sin^2\theta_{34}\ =\ \frac{\vert V_{\tau 4}\vert ^2}{1-\vert V_{e4}\vert ^2-\vert V_{\mu 4}\vert ^2},\nonumber \\
\sin^2\theta_{12}\ &=\ \frac{\vert V_{e2}\vert ^2}{1-\vert V_{e4}\vert ^2-\vert V_{e3}\vert ^2}, \ \ 
\sin^2\theta_{13}\ =\ \frac{\vert V_{e3}\vert ^2}{1-\vert V_{e4}\vert ^2},
\end{align}
\begin{eqnarray}
 \text{and } \ \sin^2\theta_{23}\ =&\ \frac{\vert V_{\mu 3}\vert ^2(1-\vert V_{e4}\vert ^2)-\vert V_{e4}\vert ^2\vert V_{\mu 4}\vert ^2}{1-\vert V_{e4}\vert ^2-\vert V_{\mu 4}\vert ^2}+ \frac{\vert V_{e1}V_{\mu 1}+V_{e2}V_{\mu 2}\vert ^2(1-\vert V_{e4}\vert ^2)}{(1-\vert V_{e4}\vert ^2-\vert V_{e3}\vert ^2)(1-\vert V_{e4}\vert ^2-\vert V_{\mu 4}\vert ^2)}.\nonumber 
\end{eqnarray}
The Dirac CP-violating phase $\delta_{CP}$, is obtained through the Jarlskog invariant $J,$ where it is defined as $J = Im[V_{e1}V_{\mu 2}V^*_{e2}V^*_{\mu 1}]$. Thus, J takes the form  \cite{KUMAR2020115082}
\begin{equation}
J = J_3 c_{14}^2c_{24}^2 + s_{24}s_{14}c_{24}c_{23}c^2_{14}c^3_{13}c_{12}s_{12}\sin(\delta_{14}-\delta_{24}),
\label{C5deltasolve}
\end{equation}
where $J_3 = s_{23}c_{23}s_{12}c_{12}s_{13}c_{13}^2 \sin \delta_{CP}$ is the Jarlskog invariant in the three neutrino framework and $s_{ij} = \sin\theta_{ij},c_{ij}=\cos\theta_{ij}$ are the neutrino mixing angles. Similarly, the two physical Majorana phases $\alpha$ and $\beta$ are determined from $V$ using the invariants $I_1$ and $I_2$ defined as follows 
\begin{eqnarray}\label{C5majoranaphase1}
I_1 = Im[V_{e1}^*V_{e2}]\ =\ c_{12} c_{13}^2 c_{14}^2 s_{12} \sin\left(\frac{\alpha}{2}\right), \\ 
I_2 = Im[V_{e1}^*V_{e3}]\ =\ c_{12}c_{13}c_{14}^2s_{13}\sin\left(\frac{\beta}{2}-\delta_{CP}\right).
\label{C5majoranaphase2}
\end{eqnarray}
To perform the numerical analysis we have fixed the values of the constant parameters as $\langle H\rangle = 246\ \text{GeV}, v = 10^{13}\text{ GeV}, \Lambda = 10^{14}\text{ GeV}, v_{\zeta} = 10^2 \text{ GeV}, M = 10^{14}\text{ GeV}$. In the active neutrino sector, there are three dimensionless free parameters $y_i$, ( $i = 1,2,3$) which are complex numbers. The real and imaginary parts of these parameters and the real coefficient $m_o$ are randomly scanned in the following ranges,
\begin{align}
\text{Re}(y_i) = [-1,1],\ \ \text{Im}[y_{i}] = [-1,1],\ \text{and}\ m_o = [0.6, 1]\ \text{eV}
\end{align}
\begin{figure}
\centering
\subfigure[]{
    \includegraphics[width=0.455\textwidth]{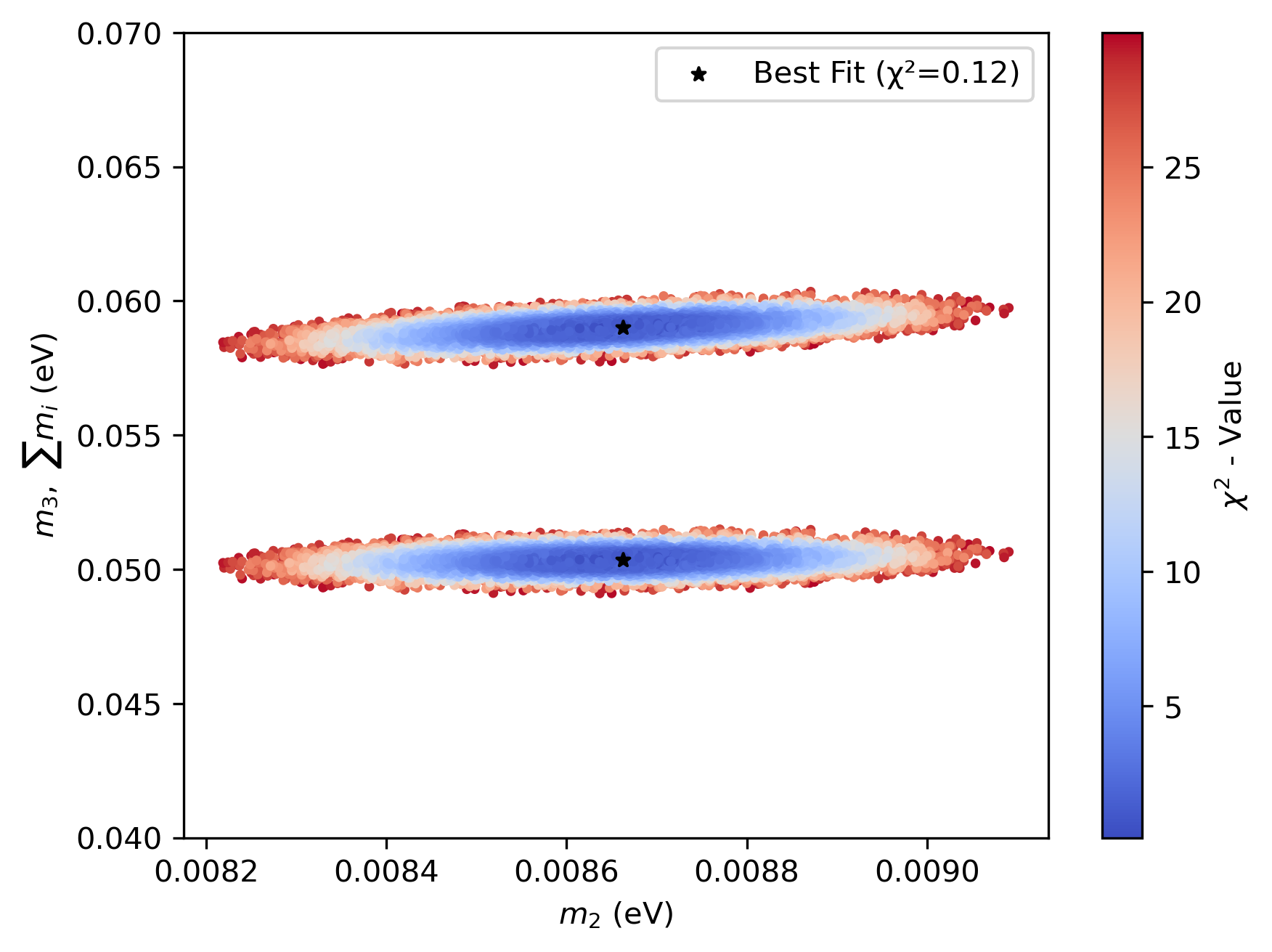}}
  \quad
\subfigure[]{
    \includegraphics[width=0.45\textwidth]{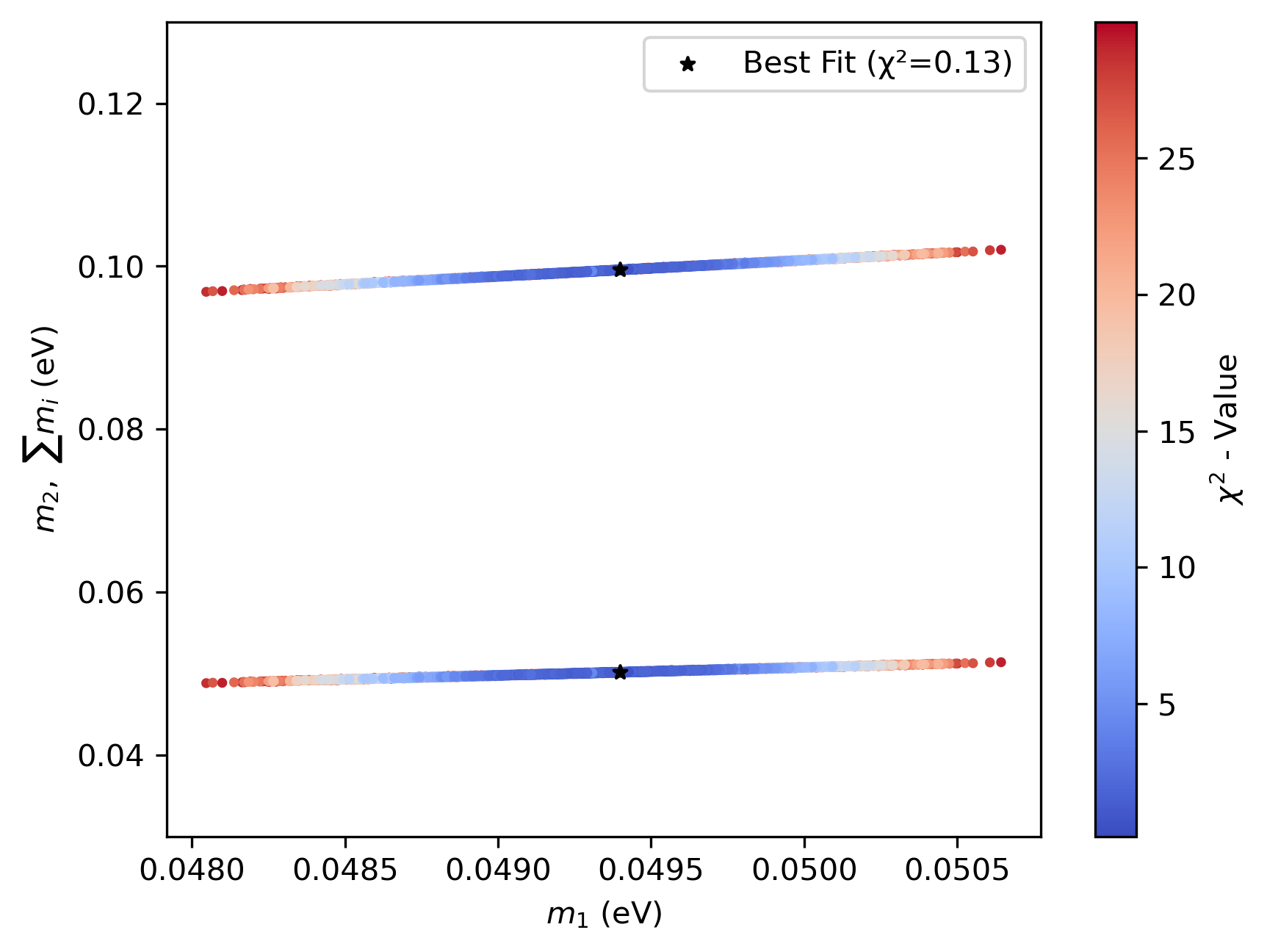}}
    \quad
  %
  \caption{\footnotesize (a). Predicted values of active neutrino masses $m_2, m_3$ and sum of active neutrino masses $\sum m_{i}$ for NH. (b) Variation of $m_{2} \text{and} \sum m_{i}$ with $m_{1}$ for IH.  }
  \label{massplot}
\end{figure}    
To fit the model parameters with the latest neutrino oscillation data \cite{esteban2025nufit}, we define a $\chi^2$ function and perform a numerical simulation using the sampling package \textbf{Multinest} \cite{feroz2009multinest}. The $\chi^2$ function is defined as
\begin{equation}
\chi^2(x_i) = \sum_{j}\left(\frac{y_j(x_i)-y_j^{BF}}{\rho_j}\right)^2.
\label{chitest}
\end{equation}
In this equation, $y_j(x_i)$ denotes the model predictions for the observables, and $y_j^{BF}$ are their best-fit values from the global analysis \cite{esteban2025nufit}. The index $j$ is summed over the neutrino observables $\left[\sin^2\theta_{12}, \sin^2\theta_{13},\sin^2\theta_{23}, \Delta m_{21}^2, \vert\Delta m_{3\ell}^2\vert\right]$, where $\ell = 1(2)$ for NH(IH) and $x_i$ are the free parameters in the model. The parameter $\rho_j$ denotes the corresponding uncertainties obtained using the $3\sigma$ range of the neutrino observables given in Table \ref{nufitdata}. However, in our analysis, the Dirac CP-violating phase $\delta_{CP}$ is not considered as an input observable, and its values are obtained as predictions from the model. By minimizing the $\chi^2$ function, we calculate the best-fit values of the model parameters and predict the best-fit neutrino observables. We have also defined the ratio of the mass-squared differences as $r= \sqrt{\Delta m_{21}^2/\Delta m_{31}^2}= m_2/m_3$ for NH $(0=m_1\ll m_2 < m_3)$ and $ r =\sqrt{\Delta m_{21}^2 /\vert\Delta m_{32}^2\vert}= \sqrt{1-m_1^2/m_2^2}$ for IH $(0=m_3\ll  m_1 < m_2)$. The results of these analyses are discussed in the next section. 
\section{Results}
\label{result}
\begin{figure}
\centering
\subfigure[]{
    \includegraphics[width=0.405\textwidth]{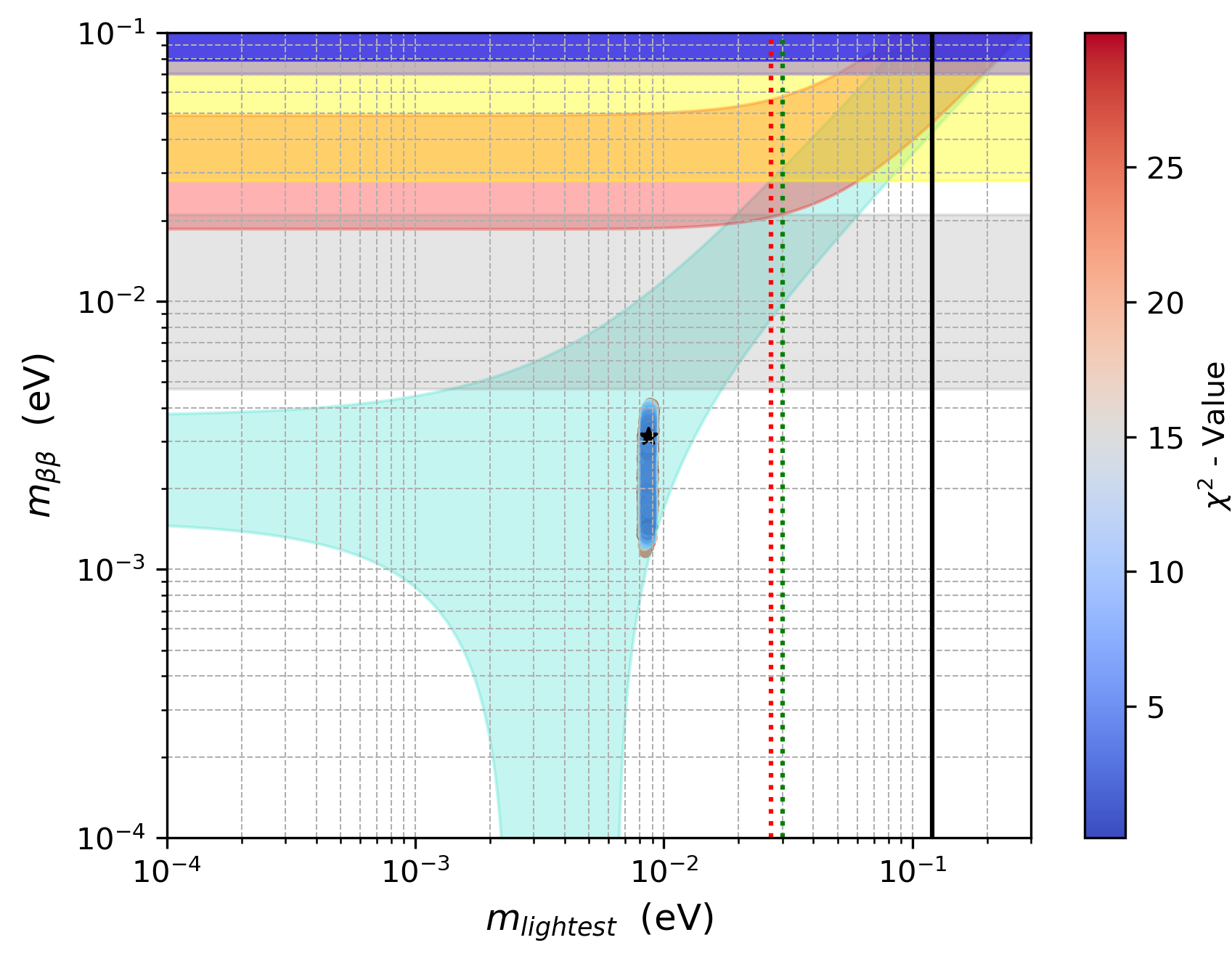}}
  \quad
\subfigure[]{
    \includegraphics[width=0.535\textwidth]{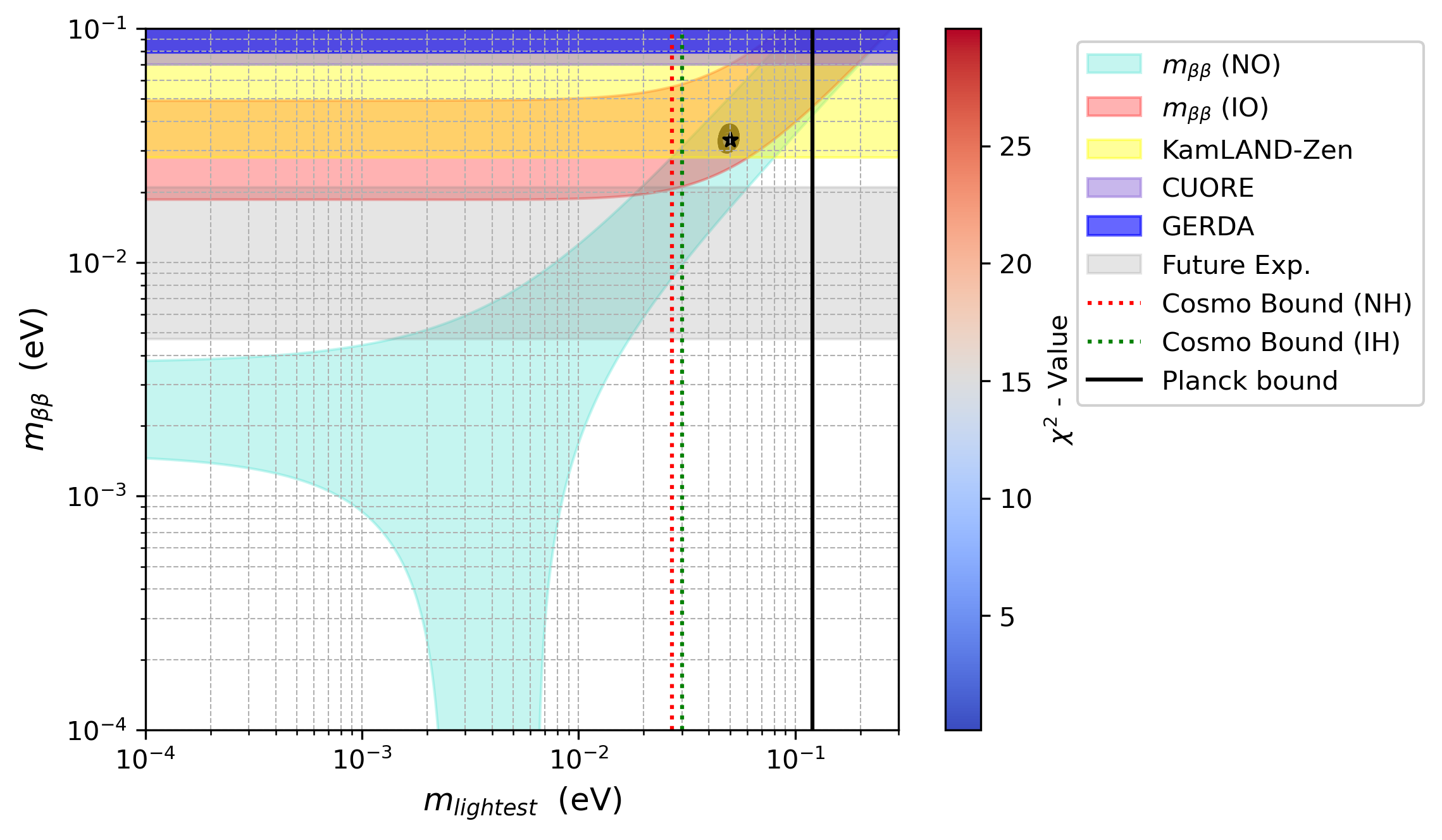}}
    \quad
  \caption{The parameter space of neutrinoless double beta decay for three neutrino mixing for NH in (a) and IH in (b). The black symbol $\bf{*}$ represents the best-fit prediction from the model. Here, $m_2 (m_1)$ is the lightest neutrino mass $m_{lightest}$ in NH (IH). The vertical black line is the Planck cosmological upper bound on sum of neutrino masses $\sum m_{i} < 0.12$ eV.}
  \label{mbbplot}
\end{figure}

In this section, we present the results of numerical analysis. The neutrino observables predicted by the model are consistent with the experimental 3$\sigma$ range given in Table \ref{nufitdata}. In these figures, the black symbol $\bf{*}$ denotes the best-fit data point of the parameters. Figure \ref{massplot} shows the predicted values of active neutrino masses for NH and IH. The best-fit values predicted by the model corresponds to $\chi^2_{min} = 0.12 $ for NH and $\chi^2_{min} = 0.13$ for IH. The best-fit model predictions and the 3$\sigma$ ranges are summarised in Table \ref{bestfit}. Here, we can observe that the model prefers a higher octant of atmospheric mixing angle $\theta_{23} > 0.50.$ The sum of the active neutrino masses are found to be $\sum m_{i} = 59.016\ \text{meV}\ (99.548\ \text{meV})$ for NH (IH), which are consistent with the latest Cosmological upper bound of $\sum m_{i}< 0.113$ eV (0.145 eV) for NH (IH). These bounds also put an upper limit on the lightest neutrino mass at $m_{lightest} \lesssim 0.027$ (0.030) eV for NH (IH). These upper bounds are shown as vertical dotted lines in Figure \ref{mbbplot}.
\begin{figure}
\centering
\subfigure[]{
    \includegraphics[width=0.45\textwidth]{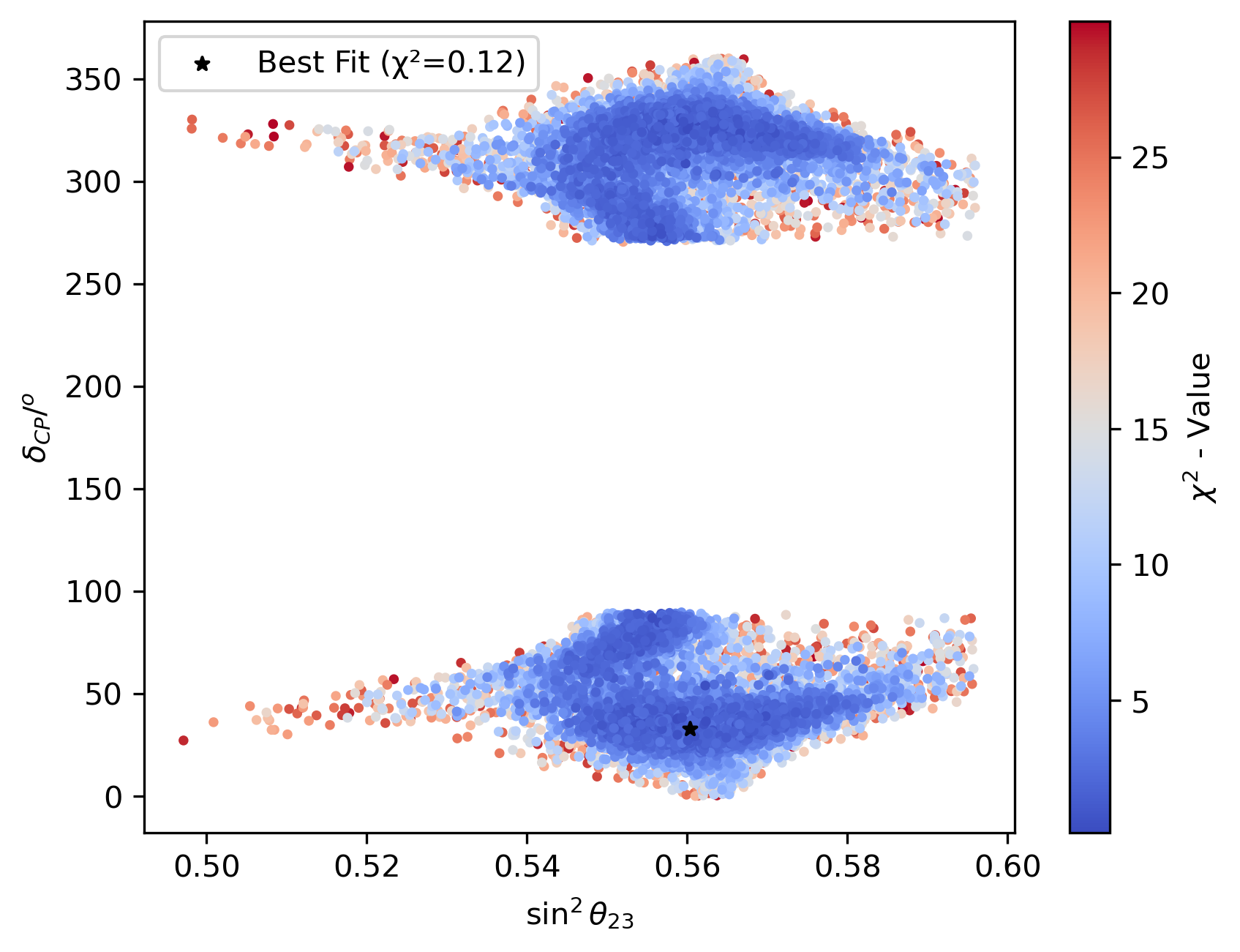}}
  \quad
\subfigure[]{
    \includegraphics[width=0.45\textwidth]{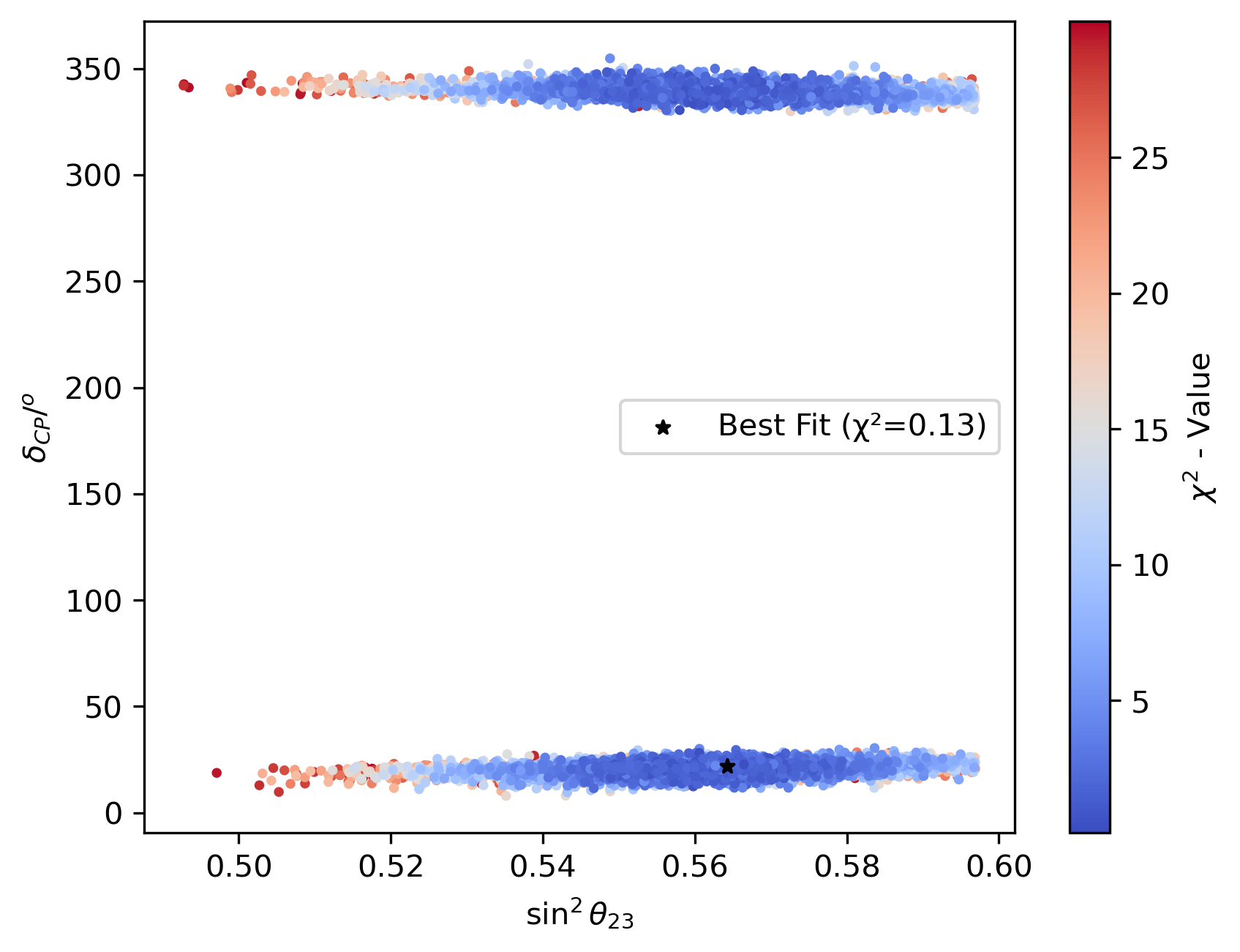}}
    \quad
  %
  \caption{Figure shows the correlation between Dirac CP-violating phase $\delta_{CP}$ with $\sin^2\theta_{23}$ for NH in (a) and IH in (b), respectively.   }
  \label{deltaphaseplot}
\end{figure} 
\begin{figure}
\centering
\subfigure[]{
    \includegraphics[width=0.45\textwidth]{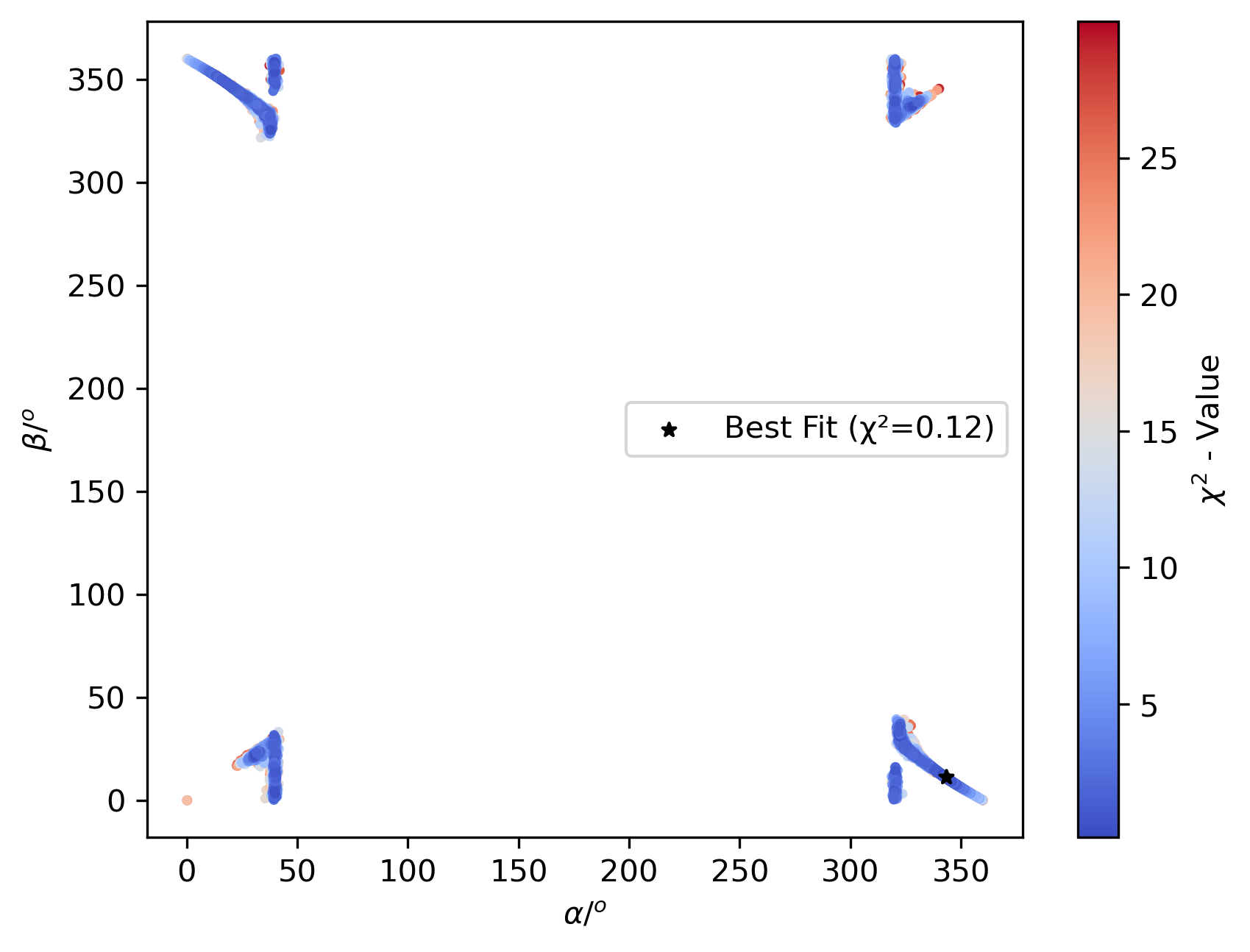}}
  \quad
\subfigure[]{
    \includegraphics[width=0.45\textwidth]{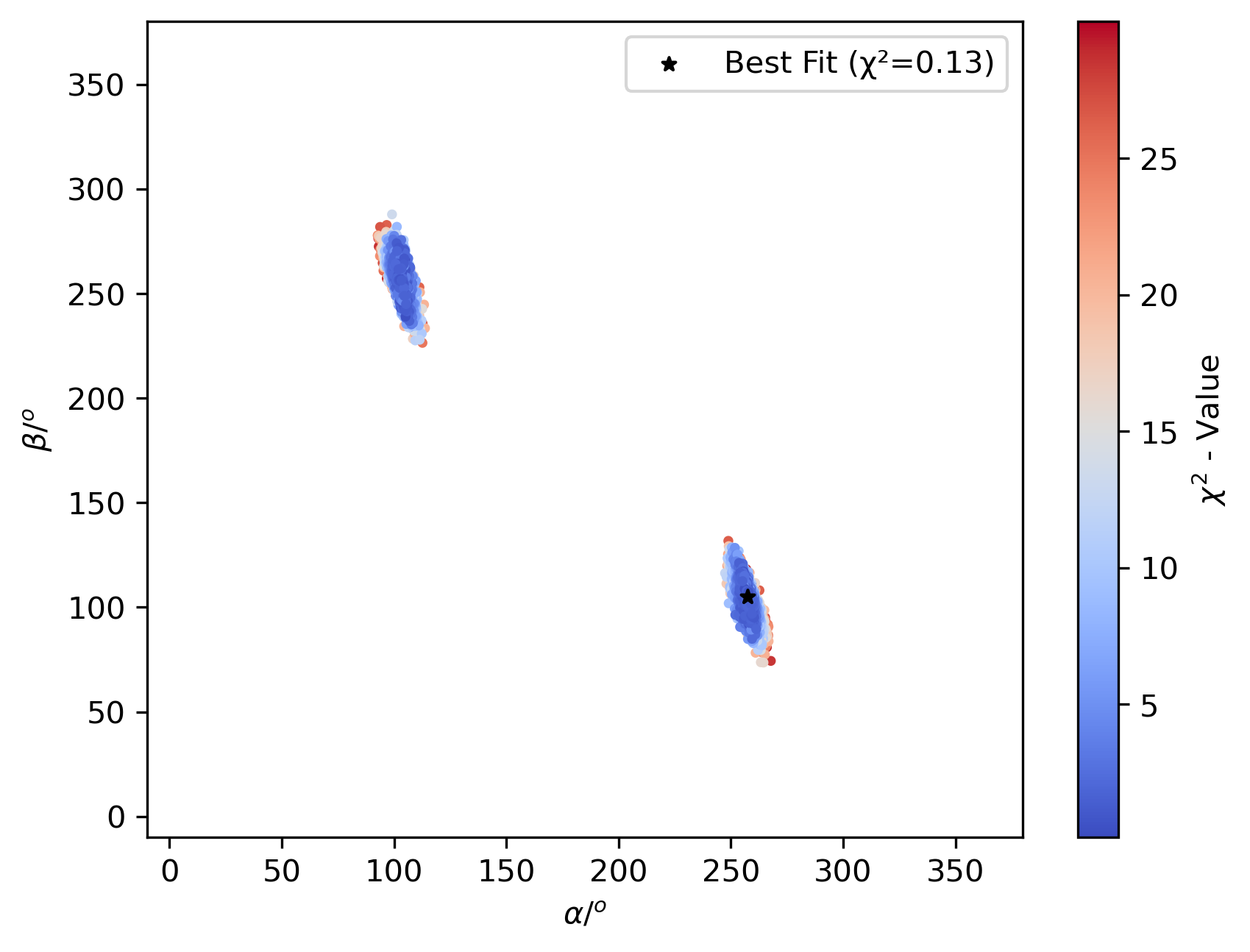}}
    \quad
  \caption{ Predicted values of Majorana phases $\alpha$ and $\beta$ for NH in (a) and IH in (b), respectively.   }
  \label{phaseplot}
\end{figure}     
Further, the effective neutrino mass in neutrinoless double beta decay is given by 
\begin{eqnarray}
m_{\beta\beta} = \vert \sum_{i=1}^3 U_{ei}^2 m_i \vert = \left\{
\begin{array}{l}
\vert m_2 s_{12}^2 c_{13}^2 e^{i\alpha} + m_3 s_{13}^2 e^{i \beta}\vert\hspace{0.15 cm} \mbox{for NH,}\ \  \\
\vert m_1 c_{12}^2 c_{13}^2 + m_2 s_{12}^2 c_{13}^2 e^{i \alpha}\vert
\hspace{0.15 cm} \mbox{for IH.}
\end{array}
\right.
\end{eqnarray}
In the 3+1 neutrino scenario, the $3\times 3$ mixing matrix $U$ is replaced by $V$, and an additional term $\vert V_{14}^2  m_4\vert$ is added to the above equation. Currently, the most stringent limit on the neutrinoless double beta decay half-life is provided by the KamLAND-Zen experiment \cite{abe2024search}. Results from this experiment constrain the effective mass parameter to $m_{\beta\beta} < (0.028 - 0.122)$eV, along with the uncertainties of nuclear matrix element (NME). Other experiments such as CUORE \cite{cuore2022search} and GERDA \cite{agostini2020final} puts an upper limit at $m_{\beta\beta} < (0.070-0.240)$eV and $m_{\beta\beta}<(0.079-0.180)$eV, respectively. There are other future experiments like nEXO \cite{adhikari2021nexo}, LEGEND-1000 \cite{abgrall2021legend} and CUPID \cite{alfonso2023cupid} which aim to increase the sensitivity to $m_{\beta\beta} \sim (0.0047 - 0.021)$eV.   
\begin{table}
\centering
{\begin{tabular}{@{}c|cccc @{}} 
\hline
\rule{0pt}{3ex} Observables & \multicolumn{2}{c|}{Best-fit} &  \multicolumn{2}{c} {3$\sigma$ range} \\
\cline{2-5}
& \rule{0pt}{3ex} NH & IH & NH & IH \\
\hline
\rule{0pt}{3ex}$\sin^2\theta_{23} $ & 0.560  & 0.564 & - & - \\ 
$\sin^2\theta_{13}$ & 0.021 & 0.022 & - & -  \rule{0pt}{3ex}\\ 
$\sin^2\theta_{12}$ & 0.307  & 0.309 & - & - \rule{0pt}{3ex}\\ 
$r$ & 0.172 & 0.172 & [0.163,0.180] & [0.160,0.184] \rule{0pt}{3ex} \\ 
$\delta_{CP}/^{o}$ & 0.573 & 0.381 & [-1.563, 1.564] & [-0.526, 0.531]\rule{0pt}{3ex} \\
$\alpha /^{o}$ & 343.28 & 257.23 & [0.03, 360] & [92.72, 267.78] \rule{0pt}{3ex}\\
$\beta /^{o}$ & 11.17  & 104.99 & [0 , 360] & [73.57 , 287.83]\rule{0pt}{3ex} \\
$m_1/$ meV & 0  & 49.39 & 0 & [48.04 , 50.64]  \rule{0pt}{3ex}\\
$m_2/$meV & 8.66 & 50.14 & [8.21 , 9.09] & [48.88 , 51.37]   \rule{0pt}{3ex}\\
$m_3/$meV & 50.35 & 0 & [49.10 , 51.49] & 0   \rule{0pt}{3ex}\\
$m_4/$ eV & 4.26 & 6.168 &[1 , 10]  & [1 , 10]  \rule{0pt}{3ex}\\
$\sum$ $m_i $/meV & 59.01   & 99.54 & [57.63,60.34] & [96.85, 102.02] \rule{0pt}{3ex} \\
$m_{\beta\beta}/$meV & 3.135  & 33.52 & [1.16, 4.13] & [31.76 , 36.06] \rule{0pt}{3ex} \\
\hline 
\rule{0pt}{3ex}
 Parameters & \multicolumn{2}{c|}{Best-fit} & \multicolumn{2}{c}{3$\sigma$ range} \\
\cline{2-5}
& \rule{0pt}{3ex} NH & IH & NH & IH \\
\hline 
Re$[y_1]$ & -0.029 & $0.033$ & [-0.052,0.049] & [-0.0401 , 0.0400] \rule{0pt}{3ex} \\ 
Im$[y_1]$ & -0.035 & $0.033$ & [-0.049,0.050] & [-0.0406 , 0.0405] \rule{0pt}{3ex} \\
Re$[y_2]$ & 0.0012   & 0.047  & [-0.026,0.026]  & [-0.189 , 0.190]  \rule{0pt}{3ex} \\  
Im$[y_2]$ & -0.02 & $-0.157$ & [-0.027,0.026] & [-0.190,0.190] \rule{0pt}{3ex} \\ 
Re$[y_3]$ & 0.105& $-0.085$ & [-0.147, 0.147] &  [-0.2604 , 0.2610] \rule{0pt}{3ex} \\
Im$[y_3]$ & 0.098 & 0.208 & [-0.147, 0.146]  & [-0.2608 , 0.2618] \rule{0pt}{3ex}\\ 
\hline 
\end{tabular} 
\caption{Best-fit values and 3$\sigma$ ranges of the model parameters and the corresponding predictions of neutrino observables for $\chi^2_{min} = 0.12 $ for NH and  $\chi^2_{min} = 0.13 $ for IH. Neutrino observables $\sin^2\theta_{12},\sin^2\theta_{13}\ \text{and}\ \sin^2\theta_{23}$ are constrained by 3$\sigma$ values from experimental data.}
\label{bestfit}}
\end{table} 
The numerical predictions of $m_{\beta\beta}$ from the model are shown in Figure \ref{mbbplot}(a) for NH and Figure \ref{mbbplot}(b) for IH. In the MES mechanism, $m_2 (m_1)$ is the lightest neutrino mass $m_{lightest}$ for NH (IH). It is observed that the predicted value of $m_{\beta\beta}$ lies well inside the experimental bounds shown in blue color, and it is also beyond the exclusions from current and future sensitivities of many experiments. However, in the case of IH, the predicted values are found inside the NME uncertainty regions of KamLAND-Zen and are shown as the horizontal yellow band. Thus, IH is highly constrained and can be excluded by future experiments if no significant signals are obtained. Further, $m_{lightest}$ is beyond the cosmological upper bound of $m_{lightest} < 0.030$ eV, as shown by the green vertical dashed line. Finally, sterile neutrino mixings slightly increase the predicted $m_{\beta\beta}$ values in NH and IH. 
\section{Conclusion}
\label{conclusion}
We have successfully constructed a new neutrino mass model based on $A_4$ symmetry by extending the SM with an $A_4$ triplet right-handed neutrino $\nu_R$ and a singlet sterile neutrino $S$ in the 3+1 scheme. An additional symmetry $Z_3\times Z_2$ is also imposed to avoid certain unwanted interactions in the Lagrangian. The model uses three triplet scalars and one singlet scalar. It is found to give both the normal and inverted mass hierarchies at 3$\sigma$. Deviation from $\mu-\tau$ symmetry is naturally obtained through the antisymmetric product of a triplet scalar $\phi_2$. We have conducted the numerical analysis using the 3$\sigma$ bounds of neutrino observables so that all the neutrino observables evaluated from the model simultaneously satisfy the 3$\sigma$ bounds. We have used the random sampling package Multinest to study the parameter space the model allows. Our analysis of neutrino masses is also consistent with the new cosmological upper bound on the sum of neutrino masses $\sum m_i < 0.113(0.145)$ eV for NH(IH).  The model predicts a higher octant of atmospheric neutrino mixing angle. In the neutrinoless double beta decay study, the predicted values are safe from the experimental exclusion region of various experiments. However, in the IH case, results of $m_{\beta\beta}$ is highly challenged by KamLAND-Zen data, and it is observed within the future sensitivities of upcoming experiments such as nEXO, LEGEND-1000, CUPID, etc. Thus, it is highly likely that IH will soon be accepted (or ruled out) based on the observation (non-observation) of signals in these experiments. Finally, it is important to mention that the presence of sterile neutrino mixings slightly increases the prediction of $m_{\beta\beta}$.
\bibliographystyle{unsrt} 
\bibliography{a4triplet}
\end{document}